\documentclass[twocolumn]{aa}  

\usepackage{xcolor}
\usepackage{float}
\usepackage{comment}
\usepackage{graphicx}  
\usepackage{gensymb}
\usepackage[varg]{txfonts}
\usepackage{hyperref}
\hypersetup{
    colorlinks=true,
    linkcolor=blue,
    filecolor=magenta,      
    urlcolor=cyan,
   citecolor=blue,
}
\begin{document}

  \title{Chromospheric heating and magnetic topology above the shared penumbra of a $\delta$-spot: Multi-line inversions and multi-height magnetic-field extrapolations}

   \author{M. Kriginsky\inst{1}, J. Leenaarts\inst{1}, J. de la Cruz Rodr\'\i guez\inst{1}, A. Pastor Yabar\inst{1}}

   \institute{Institute for Solar Physics, Dept. of Astronomy, Stockholm University, AlbaNova University Centre, 106 91, Stockholm, Sweden
              %\email{wuchterl@amok.ast.univie.ac.at}
         }

   \date{Received ; accepted }

% \abstract{}{}{}{}{} 
% 5 {} token are mandatory
 
  \abstract
{} % aims heading (mandatory)
{We aim to characterise the thermal, kinematic, and magnetic structure above the shared penumbra of a $\delta$-spot and to determine whether a recurrent chromospheric brightening is consistent with magnetic reconnection involving a twisted magnetic structure above the polarity inversion line.}
{We analysed spectropolarimetric observations of the \ion{Fe}{I} $617.3~\mathrm{nm}$, \ion{Ca}{II} $854.2~\mathrm{nm}$, and \ion{Ca}{II} H lines obtained with CRISP and CHROMIS at the Swedish 1-m Solar Telescope. We inferred the chromospheric atmosphere using spatially coupled non-local thermodynamic equilibrium inversions and estimated the chromospheric line-of-sight magnetic field using the weak-field approximation. The photospheric and chromospheric magnetic constraints were combined with a larger-scale magnetogram from the Helioseismic and Magnetic Imager and used as input to a multi-height neural-network force-free extrapolation. We investigated the reconstructed magnetic topology using the twist number, squashing factor, electric-current density, and magnetic-field-line connectivity.}
{The \ion{Ca}{II} H magnetic-field signal is concentrated mainly above the strongest photospheric field concentrations, while \ion{Ca}{II} $854.2~\mathrm{nm}$ yields generally stronger and more spatially extended line-of-sight fields. The chromosphere above the shared penumbra is approximately $300~\mathrm{K}$ hotter than nearby quiet regions around $\log\xi=-3.5$. The selected brightening follows an apparent chromospheric loop and is associated with enhanced temperature and a transition from blueshift to redshift along the structure. The extrapolation recovers magnetic-field strengths broadly consistent with the inversions and reveals a left-handed flux-rope-like core following the polarity inversion line. Enhanced electric currents and strong connectivity gradients occur near parts of its boundary, where field lines associated with the brightening connect the twisted structure to overarching loops.}
{The line-of-sight velocity pattern, and magnetic topology are consistent with a scenario in which reconnection between the twisted polarity-inversion-line field and the surrounding loops deposits energy in the chromosphere and drives plasma along the reconfigured field. These signatures do not uniquely establish reconnection, but demonstrate that combining high-resolution spectropolarimetric inversions with multi-height extrapolations can relate chromospheric energy release to the local three-dimensional magnetic structure. The recurrence of similar brightenings, together with a more energetic event shortly after the observations, motivates a time-dependent analysis of the twist, electric currents, connectivity gradients, and magnetic connectivity to search for a progressive build-up or reorganisation of the non-potential field.}


   \titlerunning{}
   \authorrunning{M. Kriginsky et al.}
   \maketitle
   \nolinenumbers
   
%
%-------------------------------------------------------------------

\section{Introduction}
Magnetic reconnection and the dissipation of waves are believed to play an important role in the heating of the solar atmosphere. A correct theoretical and  and observational interpretation of the energy transport dynamics of the solar plasma therefore requires a proper modelling of these two mechanisms.

\begin{figure*}
\centering
\includegraphics[width=18cm]{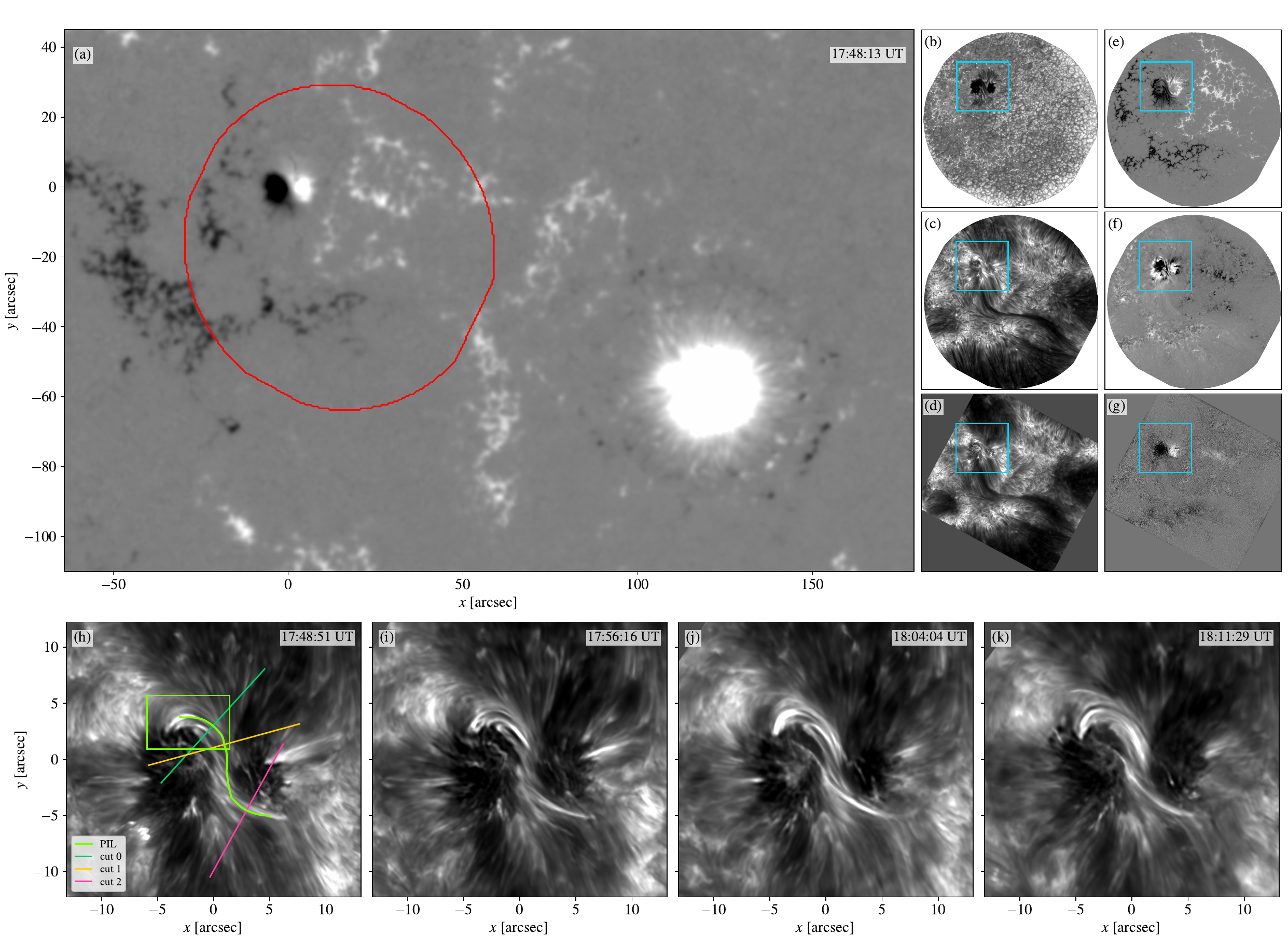}
\caption{Overview of the observations. Panel (a) shows the HMI magnetogram used for the magnetic field extrapolations, with the SST FOV outlined in red. Panels (b)--(d) show line-core intensity images of the \ion{Fe}{I} 617.3 nm, \ion{Ca}{II} 854.2 nm, and \ion{Ca}{II} H lines, respectively, at the beginning of the observing sequence. Panels (e)--(g) show the corresponding Stokes $V$ images at wavelength offsets of $-70$, $+375$, and $+130$ m\AA, respectively. The cyan boxes mark the FOV used for the inversions. Panels (h) to (k) show the integrated intensity of the \ion{Ca}{II} H line for four different times inside the cyan field of view. The green rectangle, PIL outline and cuts marked in panel (h) are references for Fig.\ref{Figure:3} and for Fig.\ref{Figure:single} and Fig.\ref{Figure:4}.}
\label{Figure:1}%
\end{figure*}
Magnetic reconnection changes the magnetic-field topology and enables stored magnetic free energy to be transferred to the plasma as heating, bulk flows, particle acceleration, and magnetohydrodynamic waves. Depending on the spatial scale and atmospheric height of the energy release, its observational manifestations range from penumbral microjets \citep[][]{katsukawa2007,tiwari2018} and Ellerman bombs \citep[][]{georgoulis2002} to UV bursts \citep[][]{peter2014,young2018}, microflares \citep[][]{hannah2011}, and major flares. These events may be identified through localized brightenings, enhanced temperatures, line broadening, strong Doppler shifts, jets, plasmoids, or bidirectional flows \citep[][]{robustini2018,diazbaso2021}, although no individual signature uniquely demonstrates that reconnection has occurred.

Energy deposited as heat in the chromosphere increases its radiative output, and the resulting radiative losses therefore constrain the energy that must be supplied to the plasma. Wavelength-integrated \ion{Ca}{II} K brightness has accordingly been used as a tracer of chromospheric radiative losses \citep[][]{leenaarts2018}. In a chromospheric reconnection event, \citet[][]{diazbaso2021} found an approximately linear relation between wavelength-integrated \ion{Ca}{II} K intensity and the radiative cooling rate integrated over atmospheric height, with a Pearson correlation coefficient of 0.97. \citet[][]{yadav2022} similarly found that the spatial distribution of height-integrated radiative losses during a C-class flare closely followed the wavelength-integrated \ion{Ca}{II} H and K emission.

Among the solar structures most favorable for energetic magnetic reconnection are $\delta$-spots, which consist of opposite-polarity umbrae embedded within a common penumbra. Their compact configuration commonly produces a strong magnetic field and a highly sheared polarity inversion line (PIL), above which substantial magnetic free energy can accumulate and subsequently be released through reconnection \citep[][]{sammis2000,toriumi2019}.

The photospheric properties of $\delta$-spots have been extensively characterized. Spectropolarimetric observations have revealed strong and predominantly horizontal magnetic fields, enhanced electric currents, and pronounced magnetic and velocity shear along the PIL \citep[][]{tanaka1991,balthasar2014,jaeggli2016}. Exceptionally strong horizontal fields, reaching approximately $5{,}500~\mathrm{G}$, have also been measured at $\delta$-spot PILs \citep[][]{lozitsky2022}, while statistical studies have characterized the enhanced emergence, rotation, and flare productivity of the participating magnetic knots \citep[][]{norton2022}.

In the chromosphere, H$\alpha$ observations have revealed the development of strongly sheared filamentary structures above the PIL \citep[][]{athay1985} and the formation of a filament during the emergence and evolution of a $\delta$-spot \citep[][]{lites1995}. \citet[][]{balthasar2014} detected \ion{Ca}{II} $854.2~\mathrm{nm}$ line-core emission and large line-of-sight velocities of both signs along the internal dividing line of a $\delta$-spot. Through inversions of \ion{Fe}{I} $630.2~\mathrm{nm}$ and \ion{Ca}{II} $854.2~\mathrm{nm}$ observations, \citet[][]{robustini2018} associated fan-shaped jets with reconnection between an emerging loop-like field and the pre-existing umbral field. Chromospheric emission during a C-class flare in a $\delta$-spot was studied by \citet[][]{guglielmino2016}, who identified \ion{Ca}{II} H flare ribbons and a Y-shaped structure extending into the corona.
%%%%%%

A physically consistent interpretation of a candidate reconnection event requires constraints on both the plasma state and the surrounding magnetic field. Multi-line non-local thermodynamic equilibrium (non-LTE) inversions provide height-dependent estimates of temperature, velocity, and magnetic field \citep[][]{delacruzrodriguez2019stic}. Their diagnostic reach is now expanding through blue and near-ultraviolet spectropolarimetry. The Visible Spectro-Polarimeter (ViSP) at the Daniel K. Inouye Solar Telescope (DKIST) \citep[][]{rimmele2020dkist,dewijn2022visp} provides full-Stokes observations down to 380 nm and has already been used for \ion{Ca}{II} H spectropolarimetry of sunspot shocks \citep[][]{french2023umbralflash} and high-resolution \ion{Ca}{II} H flare spectroscopy \citep[][]{tamburri2026caiih}. The Sunrise Ultraviolet Spectropolarimeter and Imager (SUSI) aboard the Sunrise III balloon-borne observatory extends full-Stokes measurements to 309$-$417 nm \citep[][]{feller2025susi}, enabling many-line studies across the photosphere and chromosphere \citep[][]{jafarzadeh2026multiline}. At the Swedish 1-m Solar Telescope (SST) \citep[][]{scharmer2003sst}, the recent addition of a polarimeter to the CHROMospheric Imaging Spectrometer
\citep[CHROMIS;][]{2017psio.confE..85S} enables imaging spectropolarimetry in the \ion{Ca}{II} H and K lines \citep[][]{scharmer2026spectropolarimeters}. Numerical response functions and non-LTE inversions indicate that the \ion{Ca}{II} H and K lines retain sensitivity to the upper-chromospheric line-of-sight magnetic field, particularly through Stokes $V$ in strong-field regions \citep[][]{kriginsky2026caiihk}. \citet[][]{kriginsky2026caiihk} also tested the weak-field approximation (WFA) \citep[][]{landideglinnocenti2004} using synthetic \ion{Ca}{II} H profiles. More recently, \citet[][]{juanikorena2026plage} applied the WFA to Sunrise III spectropolarimetric observations of the \ion{Ca}{II} K and \ion{Ca}{II} $854.2~\mathrm{nm}$ lines in active-region plage. They inferred chromospheric line-of-sight fields of approximately $100$--$400~\mathrm{G}$ and found that the higher-forming \ion{Ca}{II} K line produced smoother and more spatially extended magnetic-field maps than \ion{Ca}{II} $854.2~\mathrm{nm}$. The interpretation of these polarization signals nevertheless requires careful treatment of partial frequency redistribution, $J$-state interference, and the Hanle and Zeeman effects \citep[][]{juanikorena2025caii}. 

Parallel progress has been made in reconstructing the three-dimensional magnetic field. Nonlinear force-free field (NLFFF) extrapolations conventionally rely on photospheric vector magnetograms, although the photospheric field is not generally force-free \citep[][]{wiegelmann2021forcefree}. Physics-informed neural networks (PINNs) instead optimize a continuous magnetic-field representation against both the observations and the force-free and divergence-free conditions \citep[][]{jarolim2023pinn}. This framework can incorporate magnetic measurements from multiple atmospheric heights and account for their corrugated formation surfaces; adding chromospheric constraints improves the reconstructed field and its agreement with observed coronal structures \citep[][]{jarolim2024multiheight}. Combining multi-line non-LTE inversions with multi-height magnetic extrapolations therefore allows observed heating and flows to be related more directly to the currents and magnetic topology of a candidate reconnection site.

In this work, we combine two approaches that are commonly applied independently: multi-line non-LTE spectropolarimetric inversions and three-dimensional magnetic-field extrapolations. The inversions constrain the height-dependent thermodynamic, velocity, and magnetic properties of the plasma, while the multi-height extrapolations place these measurements within a three-dimensional magnetic topology. In particular, the newly available \ion{Ca}{II} H spectropolarimetry from CHROMIS extends the magnetic diagnostics towards the upper chromosphere and provides an additional chromospheric constraint alongside \ion{Ca}{II} $854.2~\mathrm{nm}$. We apply this combined framework to CRISP and CHROMIS observations obtained at the SST to characterise the plasma and magnetic structure in and above a $\delta$-spot. We focus on recurrent brightenings observed in both \ion{Ca}{II} lines and investigate one of them as a possible signature of magnetic reconnection. We examine whether its inferred temperature and velocity structure is compatible with energy deposition and field-aligned plasma flows, and whether the extrapolation recovers a physically plausible magnetic topology around the brightening while remaining consistent with the magnetic field inferred at the atmospheric heights sampled by the observations.

\begin{figure*}
\centering
\includegraphics[width=19cm]{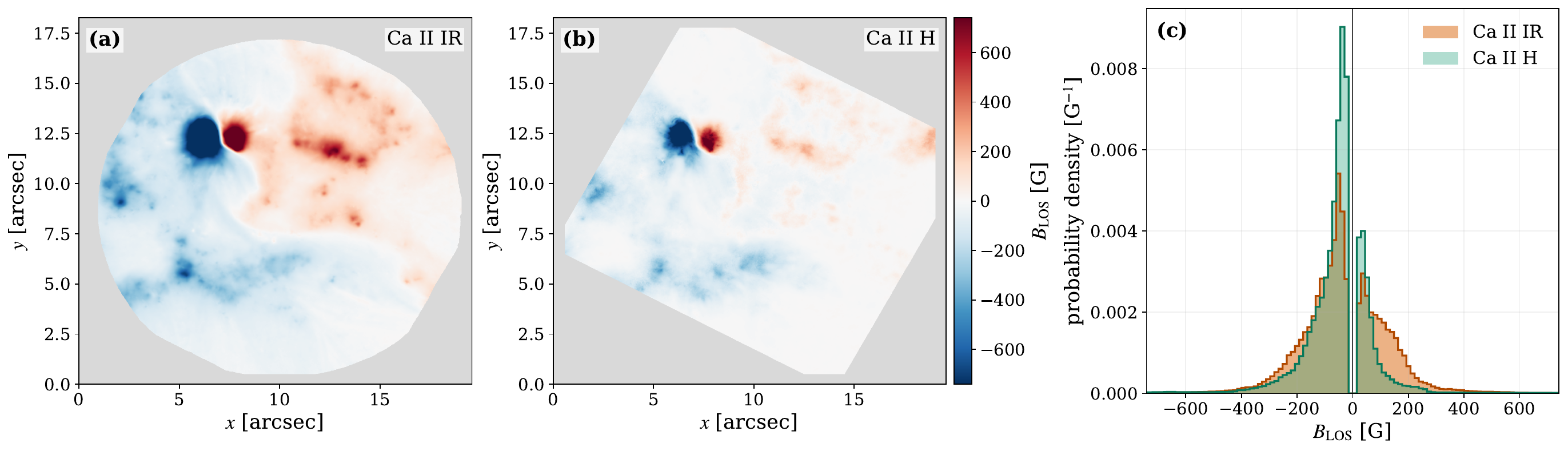}
\caption{WFA results. Panel (a) shows the WFA-obtained $B_{\mathrm{LoS}}$ for the \ion{Ca}{II} 854.2 nm line, and panel (b) shows the same quantity for the \ion{Ca}{II} H line. Panel (c) shows a histogram of their values, cropping the region around zero for readability.}
\label{Figure:2.1}%
\end{figure*}

\section{Observations} \label{subsec:observations}

The observations were obtained on 18 May 2025 with the CRISP and CHROMIS instruments at the SST. The target was the $\delta$-spot in active region NOAA 14087, located at a heliocentric angle of $\mu=0.95$. The observations covered the interval between 17:49 and 18:12 UT.

CRISP obtained full-Stokes spectropolarimetric observations of the \ion{Fe}{I} 617.3 nm and \ion{Ca}{II} 854.2 nm lines. The \ion{Fe}{I} line was sampled at 13 equidistant wavelength positions between $-24.5$ and $17.5$ pm from the line core, with a step size of 3.5 pm. The \ion{Ca}{II} 854.2 nm line was sampled at 15 positions covering the interval between $-60$ and $75$ pm from the line core. The sampling was 7.5 pm between $-30$ and $45$ pm, with a coarser sampling in the outer wings. The two lines were observed sequentially, with an average cadence of 27.7 s for the complete CRISP sequence.

CHROMIS simultaneously obtained full-Stokes spectropolarimetric observations of the \ion{Ca}{II} H line. The line was sampled at 13 equidistant wavelength positions between 396.806 and 396.884 nm, corresponding to offsets between $-39$ and $39$ pm from the line core, with a step size of 6.5 pm. An additional intensity measurement was obtained at 399.9 nm. The average cadence of the CHROMIS observations was 24.7 s. The linear polarisation signals in the \ion{Ca}{II} H line were below the noise level; Stokes $Q$ and $U$ were therefore excluded from the inversions, while Stokes $I$ and $V$ were retained. 

The data were reduced with the SSTRED processing pipeline
\citep{2015A&A...573A..40D,2021A&A...653A..68L},
including image restoration using the Multi-Object Multi-Frame Blind Deconvolution
\citep[MOMFBD;][]{2002SPIE.4792..146L,VanNoort2005}
method. The reduction also included polarimetric and intensity calibration. The \ion{Fe}{I} and \ion{Ca}{II} H observations were aligned with and resampled to the \ion{Ca}{II} 854.2 nm data. The common field of view used in the analysis covers approximately $32.5''\times30.3''$ and contains the two opposite-polarity umbrae and the interspot penumbra of the $\delta$-spot, marked on the cyan rectangles of Fig.~\ref{Figure:1}. For the curent study, the first frame of the time series was used, but frequent brightenings in the intensity of the \ion{Ca}{II} H and 854.2 nm lines were observed during the time series. Four examples are shown in panels (h) $-$ (k) of Fig.~\ref{Figure:1}. For this study, we used the first frame, shown in panel (h).

%%%%%%%%%
 
 %%%%%%%%%%
 \section{Methods} \label{sec:methods}
%%%%%%%%%%

\subsection{The inversion strategy} \label{subsec:inversion_strategy}

We used the Stockholm inversion Code
\citep[STiC;][]{2019A&A...623A..74D}
to infer the atmospheric properties within the cyan field of view marked in Fig.~\ref{Figure:1}. Given the large number of pixels  (around 
$1.20\times10^{5}$) , we first constructed a bank of representative atmospheric models that could be used to initialise the subsequent inversions.

The combined Stokes spectrum of every pixel was first decomposed using principal component analysis (PCA). We retained the scores associated with the first 35 principal components and whitened them before applying the $k$-means clustering method. A total of 50 clusters were obtained. We then identified and inverted the medoid of each cluster.

The medoids alone were insufficient to represent the spectral diversity within some of the clusters and produce an initial model guess reasonably close to the final inverted atmosphere for all pixels in the clusters. We therefore selected additional cluster members according to their distance from the profiles that had already been inverted. The distance between two profiles, $p$ and $q$, was expressed in terms of the weighted spectral $\chi^2$,

\begin{equation}
d_{\chi^2}(p,q)=
\frac{1}{N}
\sum_{j=1}^{N}
\left[
\frac{S_{p,j}-S_{q,j}}{w_j}
\right]^2,
\end{equation}
where $S_{p,j}$ and $S_{q,j}$ are the corresponding elements of the two combined Stokes spectra, $w_j$ is the weight associated with the noise of each Stokes parameter and wavelength position, and $N$ is the total number of spectral samples. Additional members were inverted until the distance between a randomly selected profile and its nearest inverted representative remained below the adopted threshold. This procedure resulted in approximately 5,000 inverted profiles and their corresponding atmospheric models, which formed the initial inversion bank.

The representative profiles were inverted in successive cycles, increasing the number of nodes used to describe the atmospheric stratification after each cycle. The atmospheric model obtained from one cycle was used to initialise the following one. Before nodes were assigned to the magnetic field components in the inversion, the WFA was applied to the observed \ion{Ca}{II} 854.2 nm line to provide an initial estimate of the magnetic field. The spatially regularised method of \citep{2024A&A...685A..85D} was applied. These estimates were used only to initialise the inversions; the final magnetic field was obtained from the full Stokes inversions with STiC.

Having obtained the bank of 5,000 initial models, we used it to provide initial guesses of the atmospheric stratification of  the different physical parameters with a spatially-coupled version of STiC. In contrast to the representative-profile inversions, for which the number of nodes was progressively increased between cycles, the coupled inversions used a fixed node distribution. We employed 12 nodes in temperature $T$, five in line-of-sight velocity $v_{\mathrm{LoS}}$, five in microturbulent velocity $v_{\mathrm{turb}}$, and two each in the line-of-sight magnetic field $B_{\mathrm{LoS}}$, horizontal magnetic field $B_{\mathrm{PoS}}$, and magnetic-field azimuth $\phi$. The patches were selected to sample the range of structures present in the observations, including quiet areas, the umbrae, and the shared penumbra of the $\delta$-spot. The atmospheric model used to initialise each pixel was selected from the representative inversion bank according to the similarity of its observed spectrum.

\begin{figure*}
\centering
\includegraphics[width=19cm]{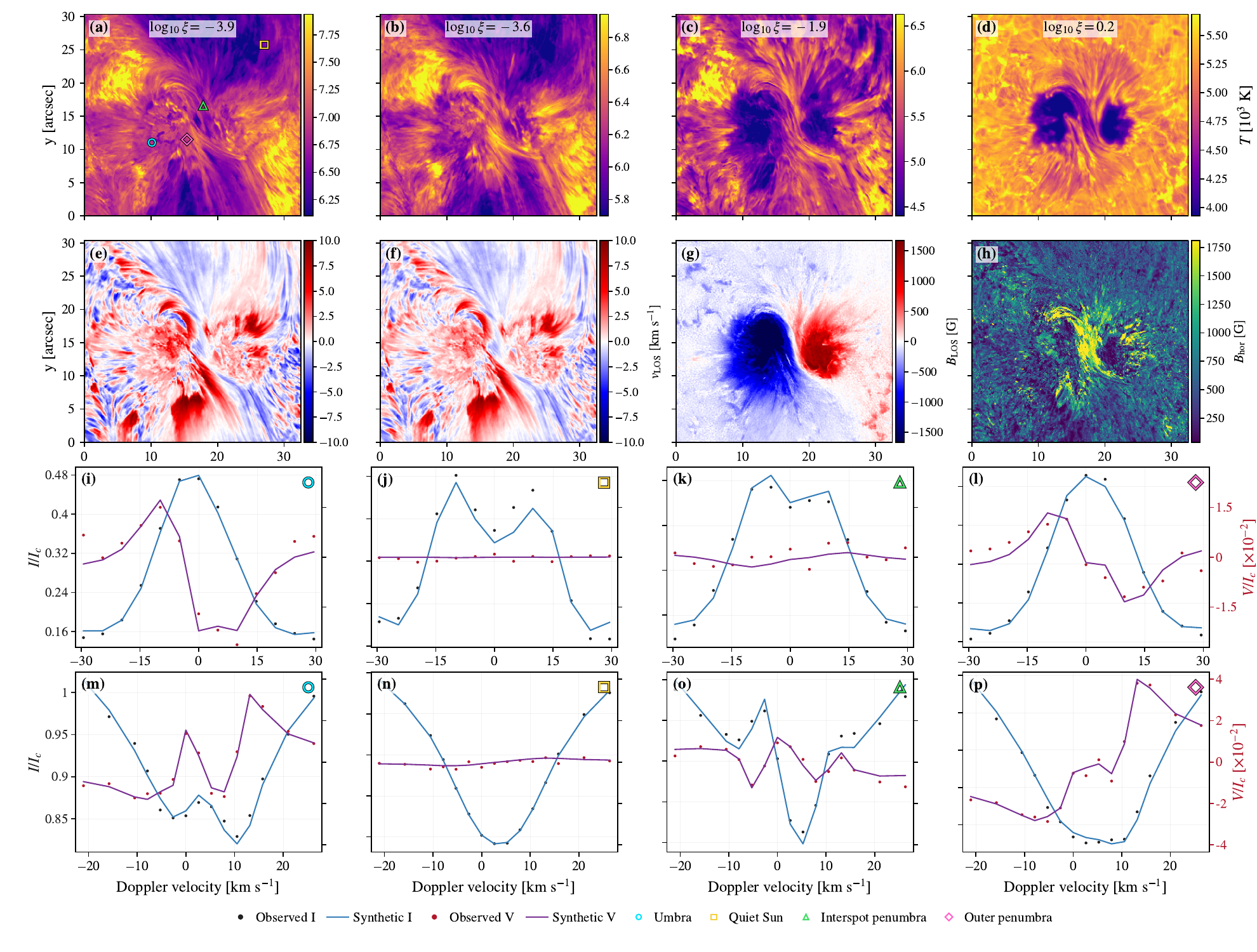}
\caption{Inversion results. Panels (a)--(d) show the temperature at $\log\xi=-3.9$, $-3.5$, $-1.8$, and $-0.2$, respectively. Panels (e) and (f) show the line-of-sight velocity at $\log\xi=-3.9$ and $-3.2$, while panels (g) and (h) show the longitudinal and horizontal magnetic field components. The symbols in panels (a) and (e) mark pixels located in the umbra, quiet Sun, interspot penumbra, and outer penumbra. Panels (i)--(l) and (m)--(p) compare the observed and fitted Stokes $I$ and $V$ profiles of the \ion{Ca}{II} H and \ion{Ca}{II} 854.2 nm lines, respectively, for the corresponding pixels.}
\label{Figure:2}%
\end{figure*}
%The coupled inversions were combined with a neural-network inversion scheme. We first trained a teacher network to infer the atmospheric model from the noise-free synthetic profiles produced by STiC. Once trained, the parameters of the teacher were kept fixed. A denoising network was then placed before the teacher and trained to transform the observed profiles into their corresponding noise-free synthetic profiles. The combination of the two networks therefore mapped an observed Stokes spectrum onto a denoised spectrum and subsequently onto its atmospheric model.

%The training sample was progressively expanded with the results of additional coupled inversions. After each set of patches, the networks were updated and applied to pixels that had not been inverted directly. The atmospheric models predicted by the network were passed through the STiC forward engine and the resulting synthetic profiles were compared with the observations. New patches were added in regions or atmospheric regimes that were not yet adequately reproduced. This process was continued until the $\chi^2$ values obtained from the network-predicted atmospheres in the remaining pixels became comparable to those obtained from the coupled STiC inversions.

\subsection{Magnetic field extrapolations} \label{subsec:extrapolations}

The photospheric vector magnetic field was inferred from Milne--Eddington inversions of the observed \ion{Fe}{I} 617.3 nm Stokes profiles, performed with the \texttt{pyMilne} code \citep[][]{delacruzrodriguez2019}. To provide the larger-scale magnetic context required for the extrapolations, the high-resolution SST magnetogram was embedded within a vector magnetogram obtained by the Helioseismic and Magnetic Imager (HMI) \citep[][]{schou2012hmi} on board the Solar Dynamics Observatory (SDO) \citep[][]{pesnell2012sdo}. We used the HMI 90-s vector-magnetic-field series \citep[][]{sun2017hmi}, whose field of view is shown in Fig.~\ref{Figure:1}(a). %The SST sequence has a cadence of 27.7 s; consequently, the HMI magnetogram nearest in time to each SST scan was used, with individual HMI maps being associated with more than one SST frame.

The magnetic field was extrapolated using the Neural Network Force-Free code
\citep[NF2;][]{2023NatAs...7.1171J}.
NF2 represents the three-dimensional magnetic field with a physics-informed neural network constrained by the observed vector field at the lower boundary. During the optimisation, the network is trained to minimise deviations from the force-free and solenoidal conditions,

\begin{equation}
(\nabla\times\mathbf{B})\times\mathbf{B}=0
\end{equation}
and
\begin{equation}
\nabla\cdot\mathbf{B}=0,
\end{equation}
throughout the extrapolation volume, while simultaneously reproducing the magnetic field at the lower boundary.

We constructed a lower boundary in which the vector magnetic field inferred from the \ion{Fe}{I} observations replaced the corresponding region of the lower-resolution HMI magnetogram. This retained the spatial resolution of the SST measurements around the $\delta$-spot while including the surrounding magnetic flux measured by HMI.  Additionally, we used the multi-height capabilities of the code to insert the WFA-computed $B_\mathrm{LoS}$ from the \ion{Ca}{II} lines, allowing the code to fit the geometrical height $z'$ of each pixel for both spectral lines. This inclusion assumes that $B_z$ in the Cartesian coordinates used in the extrapolations is equal to  $B_\mathrm{LoS}$, which for a heliocentric angle of 0.95 is reasonable but not exact.

\begin{figure*}
\centering
\includegraphics[width=18cm]{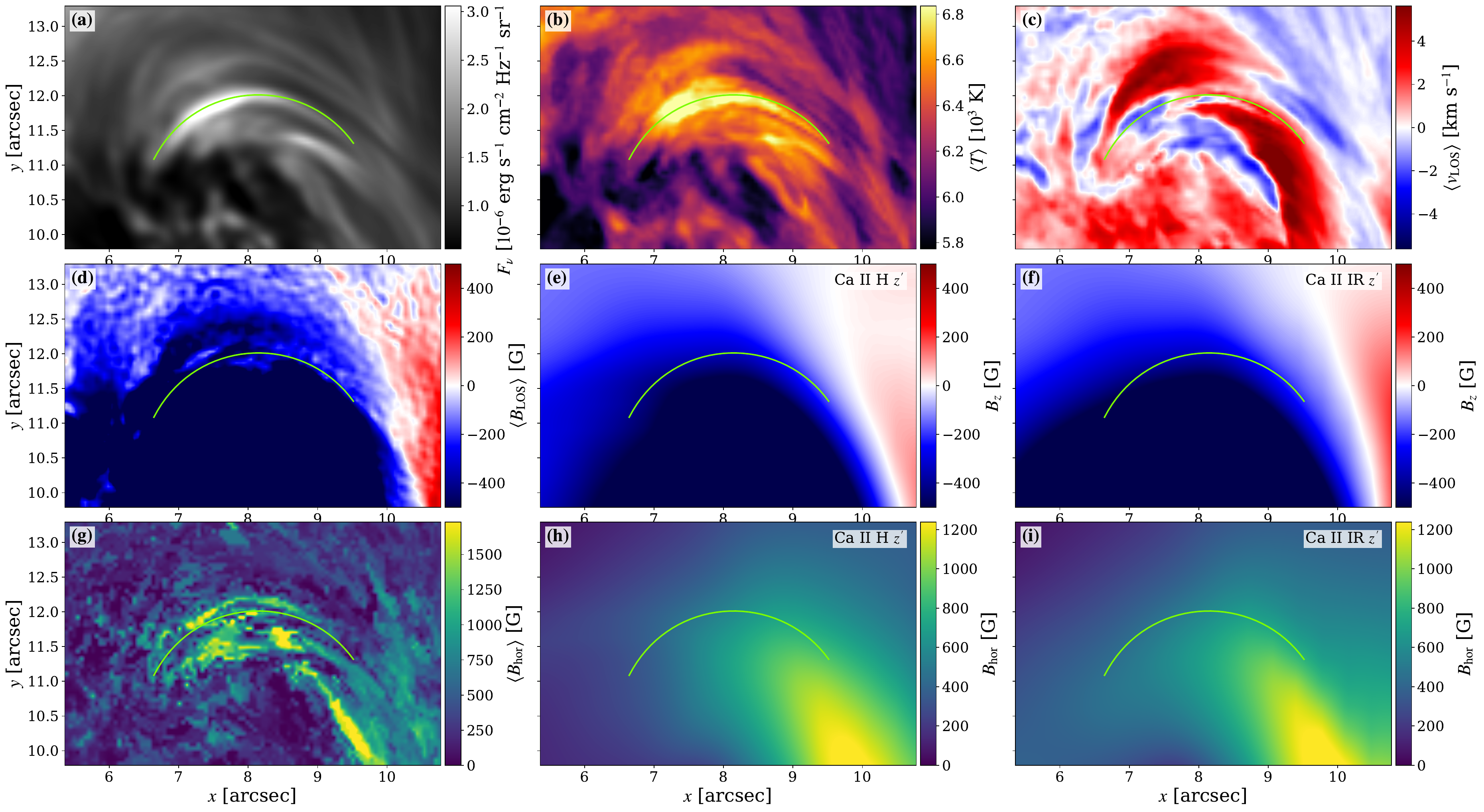}
\caption{\ion{Ca}{II} H brightening event. All panels show the region inside the green rectangle of panel (h) of Fig.\ref{Figure:1}. Panel (a) shows the integrated \ion{Ca}{II} H intensity. Panels (b), (c), (d) and (g) show the average inversion results around the $\log\xi=-3.5$ layer for $T$, $v_{\mathrm{LoS}}$, $B_{\mathrm{LoS}}$ and $B_{\mathrm{PoS}}$, respectively. Panels (e) and (h) show $B_z$ and $B_{\mathrm{hor}}$ from the field extrapolations on the inferred height of formation $z'$ for the \ion{Ca}{II} H line. Panels (f) and (i) show the same for the \ion{Ca}{II} 854.2 nm line. The green arc shown on all panels approximates the curvature of the loop family where the brightening took place.}
\label{Figure:3}%
\end{figure*}

\subsection{Magnetic topology} \label{subsec:magtop}

We analysed the magnetic topology above the PIL using the twist number, $T_{\mathrm{w}}$, and the squashing factor, $Q$, derived from the NF2 extrapolations. To define the PIL, we first computed the time-median vertical magnetic field at the lower boundary. The resulting map was smoothed over two native pixels, and the longest zero-level contour was extracted. We retained the central part of this contour that crossed the strong-gradient region between the two opposite-polarity umbrae.

We constructed a curvilinear coordinate system in which $s$ denotes the distance along the PIL, $n$ the distance normal to it, and $z$ the height above the lower boundary. The central part of the sampling path followed the magnetic PIL. Its upper extension followed a ridge identified in the \ion{Ca}{II} 854.2 nm observations, while its lower extension was obtained from a parabolic continuation of the magnetic PIL. This path was also kept fixed in time. The topology was sampled at spatial intervals of 0.1 Mm.

The field-line twist number was computed as
\begin{equation}
T_{\mathrm{w}} =
\frac{\mu_0}{4\pi}
\int_L
\frac{\mathbf{J}\cdot\mathbf{B}}{B^2},dl,
\end{equation}
where $\mathbf{J}=\nabla\times\mathbf{B}/\mu_0$ is the electric current density and the integral is evaluated along the field line $L$
\citep[e.g.][]{2006JPhA...39.8321B}.
Only complete field lines connecting two points at the lower boundary were retained. The field lines were integrated in both directions using a fourth-order Runge--Kutta scheme, trilinear interpolation of the magnetic field, and an integration step of 0.025 Mm. Tracing was terminated when the field strength fell below 1 G or after a maximum of 20,000 steps in either direction.

The squashing factor characterises the gradient of the mapping between the two lower-boundary footpoints of a field line
\citep{2002JGRA..107.1164T}.
For the mapping of a field-line footpoint $(x,y)$ onto its conjugate footpoint $(X,Y)$, the squashing factor is given by
\begin{equation}
Q = \frac{a^2+b^2+c^2+d^2}{\left|ad-bc\right|},
\end{equation}
where $a=\partial X/\partial x$, $b=\partial X/\partial y$, $c=\partial Y/\partial x$, and $d=\partial Y/\partial y$ describe the spatial derivatives of the footpoint mapping.

\begin{figure}
\centering
\includegraphics[width=9cm]{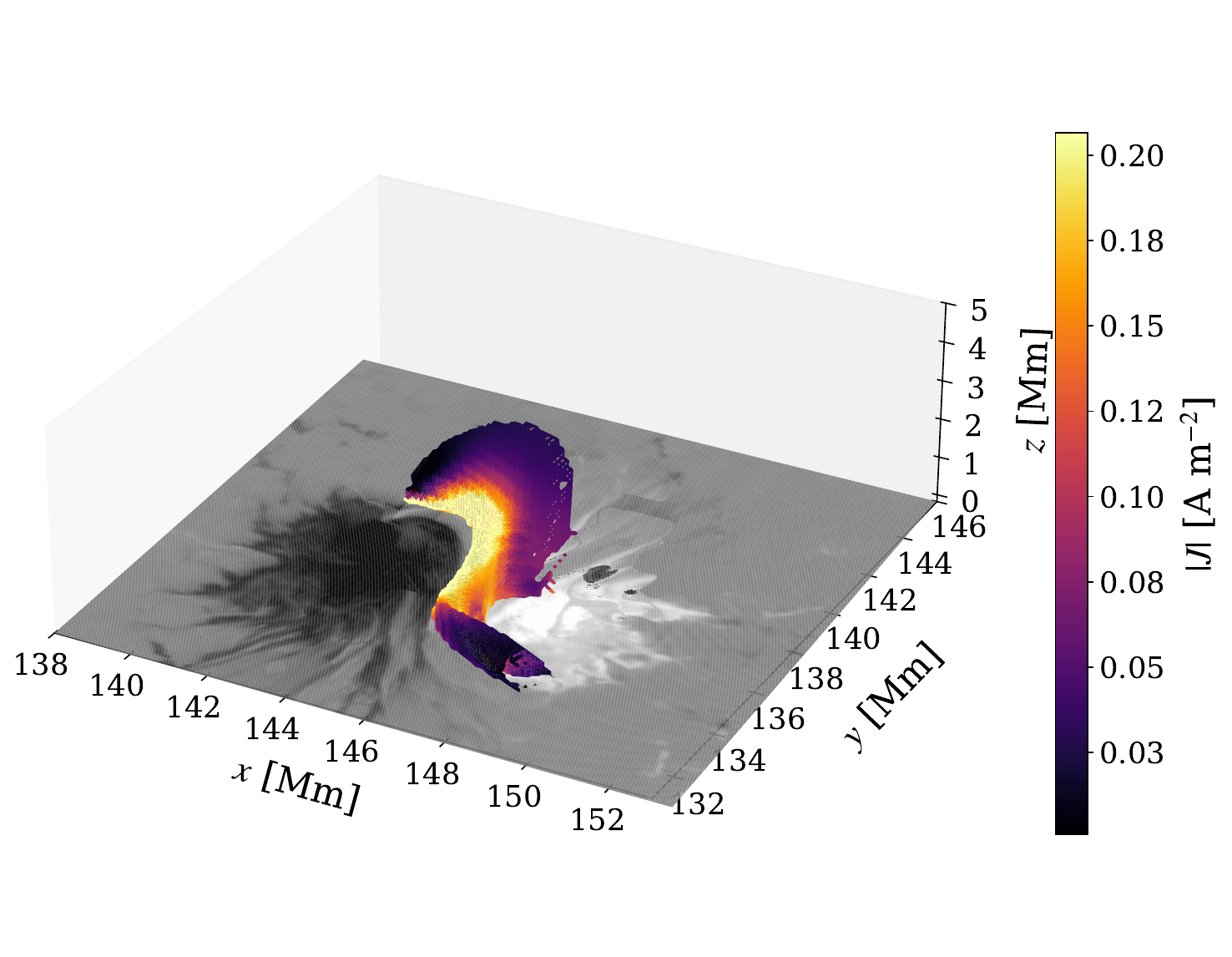}
\caption{Approximate flux-rope region. A three-dimensional contour map rendering of the high $|T_w|$ region above the PIL is shown above the photospheric $B_z$.}
\label{Figure:single}%
\end{figure}

We calculated this mapping in Cartesian coordinates at the lower boundary and transported the resulting $Q$ values along the corresponding field lines into the curvilinear volume. Singular or incomplete mappings, mappings rooted in regions of weak vertical magnetic field,  were treated as invalid. Surfaces of enhanced $Q$ were used to identify the location of strong changes in magnetic connectivity and possible quasi-separatrix layers.

To isolate the strongly twisted structure above the PIL, we first identified the largest three-dimensional connected component satisfying $T_{\mathrm{w}}\leq-0.8$. The final proxy was defined as the part of this connected envelope for which $T_{\mathrm{w}}\leq-1.25$. The connected components were determined directly in the three-dimensional $(s,z,n)$ array rather than by interpolating independently selected regions between consecutive cross-sections. We refer to the resulting structure as a flux-rope-like core. Enhanced-$Q$ surfaces, in particular those at $\log_{10}Q=2.25$ and 2.5, were used to describe its surrounding magnetic connectivity, but were not imposed as membership criteria. The selected volume therefore represents a twist-based proxy rather than a uniquely bounded flux rope.

The stability of the inferred twist was tested by varying the field-line integration step and perturbing the seed positions. These tests were used to distinguish the persistent twisted core from field lines whose inferred twist or connectivity was sensitive to the numerical tracing.

\begin{figure*}
\centering
\includegraphics[width=18cm]{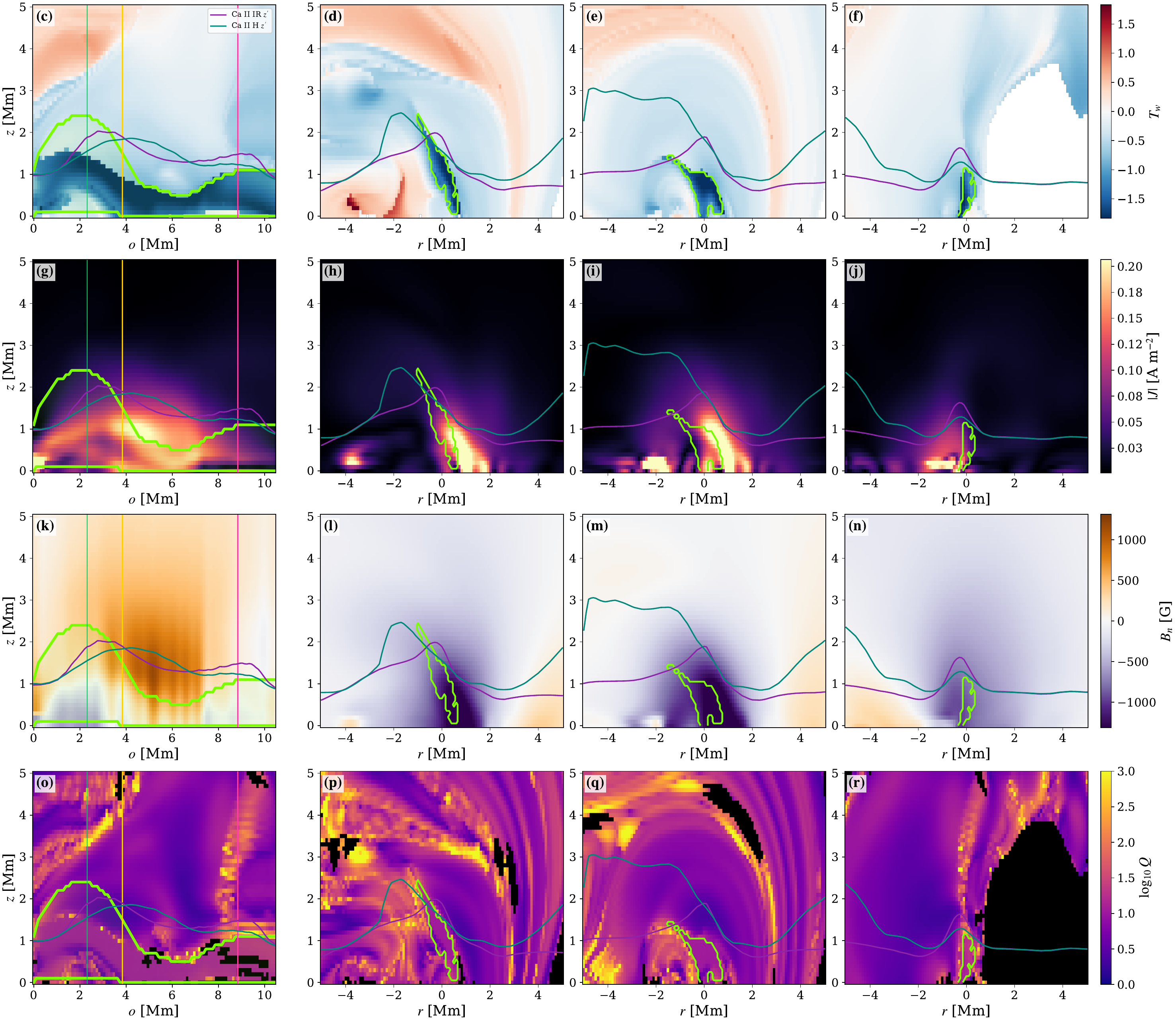}
\caption{Flux rope properties. The first row shows surface maps of $T_w$ for the path along the PIL on panel (a) and for the cross paths defined by cuts zero, one and two on panel (h) of Fig.\ref{Figure:1} on panels (b), (c) and (d). The second row show panels show, for the same surfaces, maps of $|J|$ obtained from the extrapolation. The third row shows maps of the magnetic field component perpendicular to those surfaces ($B_n$), and the bottom row shows maps of the squash factor $Q$. The green curve on the first column shows the projection into the PIL-following surface of the three-dimensional structure shown in Fig.\ref{Figure:single}. The closed green curves on the other columns outlines the cross-cut of that same structure. The formation heights of the \ion{Ca}{II} H and 854.2 nm lines as inferred by the multi-height extrapolations are shown in the dark green and purple lines, respectively.}  
\label{Figure:4}%
\end{figure*}
\section{Results}\label{sect:results}

\subsection{Weak-field approximation results}

Figure~\ref{Figure:2.1} shows the $B_{\mathrm{LoS}}$ maps obtained by applying the spatially-regularised WFA method to both \ion{Ca}{II} lines. The field inferred from \ion{Ca}{II} H is concentrated primarily above the strongest photospheric magnetic-field concentrations, with only weak values recovered elsewhere. In contrast, \ion{Ca}{II} $854.2~\mathrm{nm}$ generally yields larger $B_{\mathrm{LoS}}$ values, both above these concentrations and in relatively quiet areas, resulting in a more spatially extended magnetic-field distribution. The origin of this difference cannot be established from the observations alone. It may reflect differences in noise and magnetic sensitivity between the lines, their distinct and overlapping height sensitivities, or the weakening and lateral expansion of the magnetic field with height. These maps were used as the two chromospheric input layers for the extrapolation described in Sect.~\ref{subsec:magtop}.

\begin{figure*}
\centering
\includegraphics[width=18cm]{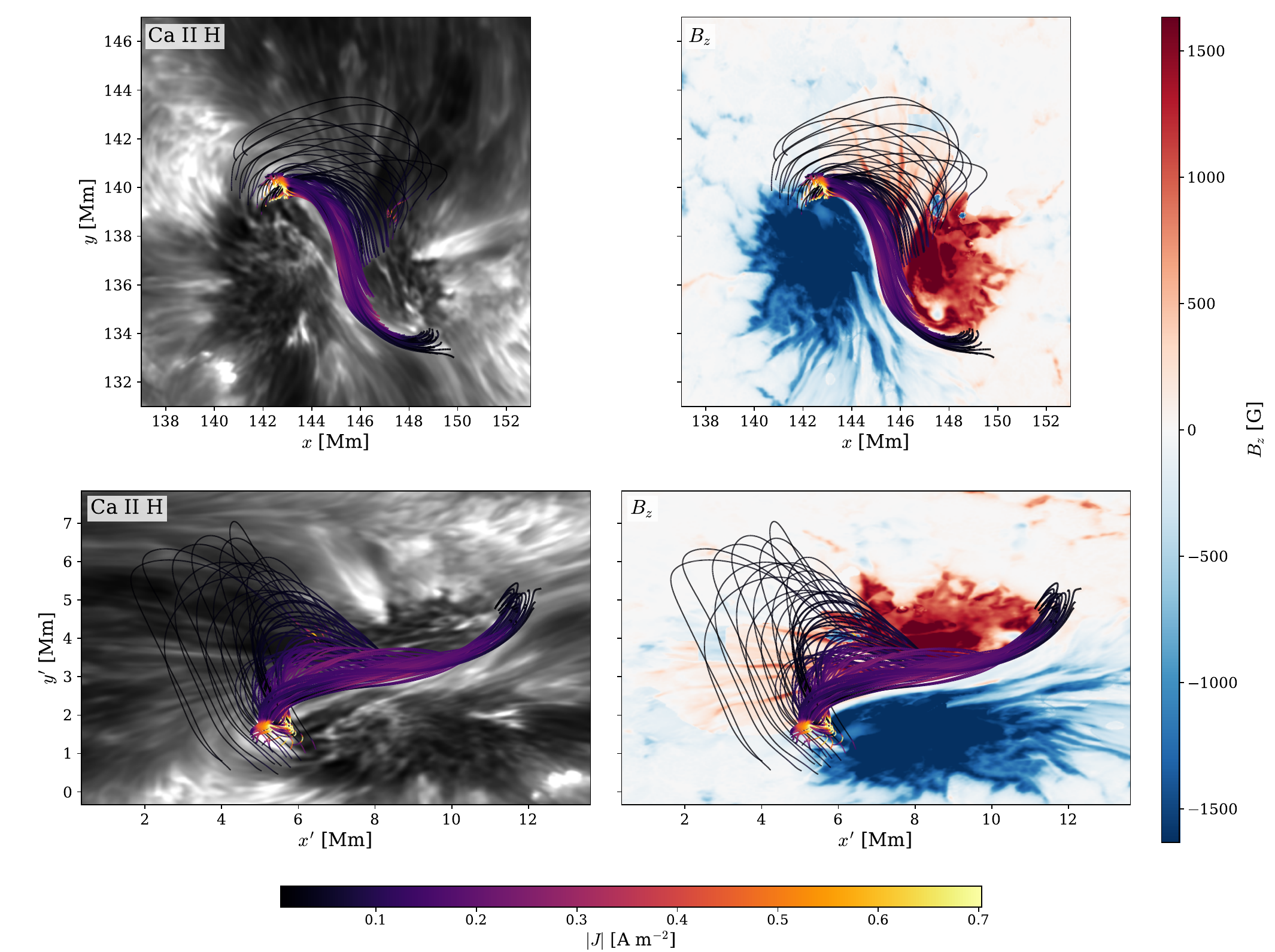}

\caption{Extrapolated field lines. Traced field lines around the location of the \ion{Ca}{II} H brightening seen in panel (a) of Fig.\ref{Figure:3}, above the intensity at the core of the \ion{Ca}{II} H (left column) and $B_z$ at the lower boundary ($z=0$) of the extrapolation result. The top and bottom rows show the same field lines, viewed from two different angles.}
\label{Figure:5}%
\end{figure*}
\subsection{Inverted atmosphere} \label{subsec:inverted_atmosphere}

The inversion strategy described in Sec.~\ref{subsec:inversion_strategy} was applied to the observations. Figure~\ref{Figure:2} shows maps of the temperature at four column-mass ($\xi$) depths, together with the line-of-sight velocity and the longitudinal and horizontal components of the magnetic field. The figure also shows examples of the observed and fitted Stokes $I$ and $V$ profiles of the \ion{Ca}{II} H and \ion{Ca}{II} 854.2 nm lines for pixels located in the umbra, quiet Sun, interspot penumbra, and outer penumbra.

The synthetic profiles generally reproduce the observed intensity and polarisation signals well. The main exception is Stokes $V$ of the \ion{Ca}{II} H line, for which the signal is dominated by noise over a substantial part of the FOV. A satisfactory fit was therefore mostly obtained in strong magnetic field concentrations with a sufficiently large line-of-sight field component. In weaker field regions, the absence of a clearly detectable Stokes $V$ signal prevented the inversions from constraining the magnetic field from this line.

The peak response to temperature closest to the line core of the two \ion{Ca}{II} lines was found to be around the $\log\xi=-3.5$ layer, where $\xi$ is expressed in g cm$^{-2}$. The intensity near the core region in both lines follow linearly the temperature at this depth.  

The mean temperature distribution at  $\log\xi=-3.5$ is on average 300 K higher in the region above the shared penumbra than in the nearby quiet regions.  We identify a region in the north side of the penumbra containing brightened loop-like structures around the core of both \ion{Ca}{II} lines. This region of interest is marked with the green rectangle in panel (h) of Fig.\ref{Figure:1}.  Panels (b), (c), (d) and (g) of Fig.\ref{Figure:3} characterise the region inside the green rectangle, with maps of $T$, $v_{\mathrm{LoS}}$, $B_{\mathrm{LoS}}$ and $B_{\mathrm{PoS}}$. All these quantities were averaged over a the range $\log\xi$ $ [-3.2, -3.7 ]$.  We identify in particular a brightening that appears to follow a family of  chromospheric loops.  The structure and curvature of the apparent loop is traced in the green path shown in all panels of Fig.\ref{Figure:3}.  The brightening region coincides with a temperature enhancement, and the trace of the loop seems to span a blue-sfhifted $v_{\mathrm{LoS}}$ region, indicating upflows which then transition into a redshifted region, representing downflows. This change in sign of $v_{\mathrm{LoS}}$ along the loop could indicate flows from one side of the loop to the other.

\subsection{The magnetic structure above the shared penumbra} \label{subsec:magnetic_structure}

Using the integrated intensity of the \ion{Ca}{II} H \& K lines as a proxy for the integrated radiative losses, we propose that the brightening might have taken place after a form of heating that deposited energy in the plasma, that at the same time caused it to flow from one side to the other of the loop structure. To test this hypothesis, we first tried to characterise the magnetic structure and topology around the $\delta$-spot.

Before interpreting the magnetic field structures around the $\delta$-spot, we tested whether the magnetic field values and heights of formation of the \ion{Ca}{II} lines estimated with the field extrapolation agreed with the inversion results. Panels (e) and (f) of Fig.\ref{Figure:3} show the maps of   $B_{\mathrm{z}}$ obtained from the extrapolation on the fitted formation height surfaces of both lines. Panels (h) and (i) of the same figure show the corresponding maps of the horizontal field component $B_{\mathrm{hor}}$.  The agreement between the inversion results and the extrapolations is reasonable, with magnetic fields of similar values being inferred using both methods, with the extrapolation results being smoother.

We then used the extrapolations to characterise the magnetic field topology in and around the PIL. Tracing the PIL at the photosphere (green path in panel (h) of Fig.~\ref{Figure:1} , we computed the $T_w$  on a surface perpendicular to the PIL at each point. We then identified a high-$T_w$ region in between the two sunspots, indicating the presence of a left-handed twisted structure. Its three-dimensional rendering is shown in Fig.\ref{Figure:single}, coloured by the value of the electric current $J$, as computed from the extrapolated field. 

Panels (b) $-$ (d) of Fig.\ref{Figure:4} show examples of the computed $T_w$ surfaces at three cut locations along the PIL, whose locations are marked on panel (h) of Fig.\ref{Figure:1}.  A central, low-$T_w$  domain was detected, surrounded by a weaker yet also negative $T_w$ region. The white region on panel (d) represents an undefined $T_w$ domain due to the lack of field lines crossing the surface determined by cut number two in that area. We also computed the $T_w$ surface tangent to the PIL, shown in panel (a), with the projection of the three dimensional structure on Fig.\ref{Figure:single} shown as the lime curves. 

The $T_w$ surfaces, along with the high-current regions near the boundary of the serpent-like high-$T_w$ structure displayed on panels (g) $-$ (j)) and the magnetic field normal to those surfaces, shown as $B_n$ on panels (k) $-$ (n) support the existence of a flux-rope like structure above the shared penumbra of the $\delta$-spot.

\subsection{Reconnection event}

The brightenings seen in the  \ion{Ca}{II} lines are recurrent over the time series, and given the presence of a flux rope-like structure in the same region, we consider whether such brightenings are produced as a result of plasma being heated as a consequence of reconnection events between the flux rope (high-$Tw$ region) and its surrounding low-$T_w$ region. 

To support this idea, we trace magnetic field lines that run across the PIL, as well as lines rooted near the location of the chromospheric loop with the increased brightening event. The results are shown in Fig.~\ref{Figure:5}. The field lines traced from the extrapolation support the possibility of a reconnection event in an area of high-$J$, which coincides with the location of the brightening viewed on the chromosphere. If the heating was driven by the reconnection, we cannot trace what the field lines might have looked like before the event, as that took place before the observations. However, the geometry of the loops that envelop the PIL-following lines make it a plausible scenario for the location of the chromospheric brightening event and the evolution of  $v_{\mathrm{LoS}}$ from blueshift to redshift as the plasma moves from one footpoint to the other.

\section{Conclusions}\label{sec}

We combined multi-line non-LTE spectropolarimetric inversions with multi-height magnetic field extrapolations to investigate the chromosphere above the shared penumbra of a $\delta$-spot. In particular, we studied a loop-like brightening observed in the \ion{Ca}{II} H and \ion{Ca}{II} $854.2~\mathrm{nm}$ lines and examined whether its plasma properties and surrounding magnetic topology were consistent with a reconnection event.

The $B_{\mathrm{LoS}}$ maps inferred with the WFA show substantial differences between the two chromospheric diagnostics. The \ion{Ca}{II} H measurements are primarily sensitive to regions above the strongest photospheric magnetic-field concentrations, whereas \ion{Ca}{II} $854.2~\mathrm{nm}$ produces stronger and more spatially extended fields. These differences may result from the distinct height sensitivities and noise properties of the lines, as well as from the weakening and lateral expansion of the magnetic field with height.

The inversions reveal that the chromosphere above the shared penumbra is, on average, approximately $300~\mathrm{K}$ hotter than the nearby quiet regions around $\log\xi=-3.5$. Within this enhanced-temperature region, the selected brightening follows an apparent chromospheric loop. The line-of-sight velocity changes from blueshift to redshift along the structure, consistent with plasma moving upwards along one part of an inclined loop and downwards beyond its apex. This interpretation remains dependent on the loop geometry because only the line-of-sight component of the velocity is measured.

The multi-height extrapolation recovers magnetic-field strengths broadly consistent with those inferred from the inversions, although the extrapolated maps are spatially smoother. The reconstructed topology contains a left-handed, strongly twisted structure following the PIL above the shared penumbra. This flux-rope-like core is surrounded by a more weakly twisted magnetic domain and contains enhanced electric currents near parts of its boundary. Field lines traced near the chromospheric brightening connect the PIL-following structure to overarching loops and pass close to a region of high current density.

Taken together, the enhanced temperature and \ion{Ca}{II} emission, the progression from blueshift to redshift, and the reconstructed magnetic topology support a scenario in which reconnection between the twisted PIL-following field and the surrounding loops deposits energy in the chromosphere and drives plasma along the reconfigured field. These signatures are nevertheless indirect and do not uniquely demonstrate that reconnection occurred, particularly because the magnetic configuration prior to the brightening is not available. We therefore identify the event as a reconnection candidate whose interpretation is consistent with both the spectropolarimetric inversions and the magnetic-field extrapolation. Extending the analysis to the full time series will be required to determine how the recurrent brightenings relate to the temporal evolution of the currents, twist, magnetic connectivity, and chromospheric energy deposition.

Similar brightenings recur throughout the SST time series, indicating that the analysed event was not isolated. A more energetic event also occurred in the same region shortly after the SST observations ended. This motivates extending the analysis to the remaining brightenings and to the complete sequence of multi-height extrapolations. Such a study could determine whether the recurrent energy release is accompanied by systematic changes in the magnitude and distribution of $T_{\mathrm{w}}$, the high-$Q$ layers, the electric-current concentrations, or the connectivity of the flux-rope-like structure. In particular, it would allow a retrospective test of whether the high-resolution SST inversions and extrapolations reveal a progressive build-up or reorganisation of the non-potential magnetic field before the subsequent event. Although such changes would not by themselves constitute a prediction, they could provide observational indicators that the magnetic configuration was evolving towards a state more favourable for a larger energy release.

\begin{acknowledgements}
The authors of this paper gratefully acknowledge funding by the European Union through the European Research Council (ERC) under the Horizon Europe program (MAGHEAT, grant agreement 101088184), and the Swedish Research Council (registration number 2022-03535 and 2021-05613), and Swedish National Space Agency (2021-00116).
The Swedish 1-m Solar Telescope is operated on the island of La Palma by the Institute for Solar Physics of Stockholm University in the Spanish Observatorio del Roque de los Muchachos of the Instituto de Astrof\'\i sica de Canarias. The Institute for Solar Physics is supported by a grant for research infrastructures of national importance from the Swedish Research Council (registration number 2021-00169). The computations were enabled by resources provided by the National Academic Infrastructure for Supercomputing in Sweden (NAISS), partially funded by the Swedish Research Council through grant agreement no. 2022-06725. The authors acknowledge the National Academic Infrastructure for Supercomputing in Sweden (NAISS), partially funded by the Swedish Research Council through grant agreement no. 2022-06725, for awarding this project access to the LUMI supercomputer, owned by the EuroHPC Joint Undertaking and hosted by CSC (Finland) and the LUMI consortium.
\end{acknowledgements}

\bibliographystyle{aa}
\bibliography{citations}

\end{document}